\documentclass[3p,times]{elsarticle}
\usepackage{amsmath}
\usepackage{bm}

\usepackage{color}
\usepackage{multirow}
\usepackage{hyperref}
\usepackage[capitalise]{cleveref} 
\usepackage{siunitx}
\usepackage{subcaption}

\newcommand{\G}{\Gamma}

\newcommand{\brho}{\bar{\rho}}
\newcommand{\qq}{{\bm q}}
\newcommand{\pp}{{\bm p}}
\newcommand{\nn}{{\bm n}}
\newcommand{\xx}{{\bm x}}
\DeclareMathOperator{\tr}{tr} 
\DeclareMathOperator{\Tr}{Tr}

\begin{document}

\begin{frontmatter}

\title{Critical behavior of isotropic systems with strong dipole-dipole interaction from the functional renormalization group}

\author{Georgii Kalagov\corref{cor1}}
\ead{kalagov@theor.jinr.ru}

\author{Nikita Lebedev}%
\ead{lebedev@theor.jinr.ru}

\address{Joint Institute for Nuclear Research, Joliot-Curie 6, 141980 Dubna, Russia
}

\cortext[cor1]{Corresponding author}

\begin{abstract}

We compute the critical exponents of three-dimensional magnets with strong dipole-dipole interactions using the functional renormalization group (FRG) within the local potential approximation including the wave function renormalization (LPA$^\prime$). The system is governed by the Aharony fixed point, which is scale-invariant but lacks conformal invariance. Our nonperturbative FRG analysis identifies this fixed point and determines its scaling behavior. The resulting critical exponents are found to be close to those of the Heisenberg $O(3)$ universality class, as computed within the same FRG/LPA$^\prime$ framework. This proximity confirms the distinct yet numerically similar nature of the two universality classes.

\end{abstract}

\end{frontmatter}

\section{Introduction}

$O(N)$-symmetric field theories provide a standard framework for analysing critical phenomena in condensed matter physics.
They describe magnetic phase transitions governed by short-range exchange interactions between $N$-component spins. In many real materials, however, long-range forces are also present and can modify critical behavior. A prominent example is the magnetic dipole-dipole interaction. It has long been hypothesized that such long-range forces could perturb  critical behavior, particularly in the exceedingly narrow vicinity of the critical point, where the correlation length becomes comparable to the scale set by the dipolar strength \cite{Kadanoff}.

The first systematic study of dipolar magnets was conducted by Fisher and Aharony \cite{fisherprl,dipolarI, Aharony2, Aharony3}. Employing Wilson’s renormalization-group (RG) approach, they studied isotropic magnetic systems in $d=4-\varepsilon$ dimensions with both short-range exchange and long-range dipolar interactions between $(N=d)$-component spins. They established that the dipolar interaction modifies the RG flow and gives rise to a distinct dipolar (often called Aharony) fixed point, whose critical exponents differ from those of the Heisenberg universality class. The corresponding exponents were obtained at leading order of the $\varepsilon$-expansion. It was also shown that the dominance of the dipolar interaction in this regime suppresses longitudinal spin fluctuations. The latter effect was confirmed experimentally \cite{SupressionExperiment}. This  justifies the strong dipole–dipole interaction model as a limiting case of the $O(N=d)$-symmetric theory for a divergence-free vector field \cite{KudlisPikelner}.

Since the original studies, considerable effort has been dedicated to verifying and improving these predictions.  The results for the critical exponents were extended to the second order of the $\varepsilon$-expansion in \cite{twoloops}. Recently, using the RG approach formulated directly in fixed spatial dimensions \cite{Baker, LeGuillou,HENRIKSSON20231},  Kudlis and Pikelner \cite{KudlisPikelner} derived analytic three-loop expressions for the critical exponents. After Borel resummation, their estimates were found to be numerically close to those of the Heisenberg universality class. This proximity suggests that dipolar effects may be subtle in practice but warrant careful consideration for precision. Despite substantial progress based on perturbative RG, diagrammatic calculations for dipolar criticality remain technically challenging, and available series are restricted to comparatively low loop orders. 
This stands in stark contrast to the standard Heisenberg $O(N)$-symmetric model, where the $\varepsilon$-expansion has attained seven loops \cite{Schnetz, Shalaby_2021, Abhignan_2021} and the RG calculations in three dimensions have achieved six loops \cite{nickel}. High-order calculations provide a benchmark for advanced resummation techniques \cite{Mera16,Mera18} and yield results (see \cite{Kompaniets2017} for a review) that agree well with other approaches such as the conformal bootstrap (CB) and Monte Carlo (MC) simulations.

In this context, it is highly desirable to obtain independent, nonperturbative estimates of the critical exponents at the dipolar critical point.  While the numerical CB can determine critical exponents of $O(N)$ models with high precision \cite{ON_bootstrap}, recent arguments indicate that the dipolar critical point is scale but not conformally invariant \cite{NakayamaRychkov2024}. The CB method is therefore not applicable in this setting, which highlights the need for approaches that do not assume conformal invariance. The functional renormalization group (FRG) is particularly well suited for this purpose. It provides a nonperturbative framework for computing the effective action and describing critical properties without relying on diagrammatic expansions in a small parameter  \cite{berges2002,pawlowski2007, dupuis2021}. The first FRG study of the dipolar fixed point was recently reported by Nakayama \cite{Nakayama2024} based on  the local potential approximation (LPA) for the effective action. A key limitation of that study is in the treatment of the anomalous dimension $\eta$: the RG flow for the effective potential was written allowing for  $\eta \neq 0$, but the wave-function renormalization was not evolved, and $\eta$ was taken from the $\varepsilon$-expansion rather than obtained from the FRG. The resulting truncation is therefore not a self-consistent FRG description of the dipolar fixed point.

In the present work, we address this gap by employing the FRG and approximating the effective action within the LPA including the wave-function renormalization, commonly referred to as the LPA$^{\prime}$ truncation. Within this framework we derive the anomalous dimension $\eta$ which closes the FRG system and enables a consistent determination of the remaining critical exponents. Our analysis is based on the effective action for a divergence-free field, which enforces the transversality constraint associated with strong dipolar dominance. By locating the fixed point of the FRG flow, we compute the critical exponents $\eta$, $\nu$ (correlation-length exponent), and $\omega$ (leading correction-to-scaling exponent), enabling a direct comparison with perturbative results.

The paper is organized as follows. In \cref{sec:model} we introduce the field-theory model of the investigated system.
In \cref{sec:frg} we outline the FRG framework, specify our nonperturbative truncation of the effective average action and derive flow equations for scale-dependent couplings.
Section~\ref{sec:results} presents our results for the critical exponents and includes a comparison between the dipolar and  Heisenberg universality classes, as well as a discussion of the agreement with earlier studies. Section~\ref{sec:conclusios} summarizes our findings and highlights possible directions for further research. Technical details are collected in the Appendices.

\section{ \label{sec:model} Effective field model }

 The critical behavior of an isotropic dipolar magnet is modeled by the continuum effective action  of a real vector field $\phi_a(\xx)$ in $d$ spatial dimensions \cite{NakayamaRychkov2024}
\begin{align}
\label{eq:action}
    S[\phi] =  \int_{\xx} \left( \frac{1}{2} (\partial_a \phi_b)^2+\frac{h}{2} (\partial_a \phi_a)^2 + \frac{u_0}{2} \phi_a\phi_a + g_0 (\phi_a\phi_a)^2 \right), \quad  \int_{\xx} \equiv \int d^dx,
\end{align}
with the usual summation convention over repeated indices,  $a, b=1,\dots,d$. The quadratic part contains the standard gradient term $(\partial_a \phi_b)^2$ and the additional longitudinal contribution $(\partial_a \phi_a)^2$ that encodes the effect of the dipole-dipole interaction at long distances. This contribution locks internal and spatial rotations to  single $O(d)$ symmetry. In the limit $h\to 0$, the model continuously approaches the Heisenberg one. The quartic self-interaction $g_0>0$ is responsible for nontrivial criticality. The parameter $u_0$ serves as a tuning variable. For a  fixed $g_0$, criticality is reached  at some value $u_0=u_{0, c}(g_0)$ corresponding to the phase transition temperature $T_c$. At this point, the system becomes scale invariant, and the correlation functions exhibit power-law behavior at large scales.  

The dipolar/Aharony universality class emerges when longitudinal fluctuations are strongly suppressed. In the present formulation, this corresponds to the limit $h\to \infty$ in which the action imposes the transversality constraint $\partial_a \phi_a=0$. In this case, only transverse modes contribute to infrared behavior, and at criticality the Fourier transform of the two-point function takes the scaling form
\begin{align}
\label{eq:2pt}
    \langle \phi_a(\qq) \phi_b(-\qq) \rangle = \frac{P_{ab}(\qq)}{q^{2-\eta_D}}, \quad P_{ab}(\qq) = \delta_{ab} -\frac{q_a q_b}{q^2}, \quad q=|\qq|,
\end{align}
with $\eta_D$ being the anomalous dimension at the dipolar fixed point. The transverse projector $P_{ab}(\qq)$ is scale invariant but its momentum dependence is not consistent with conformal covariance \cite{Nakayama2011, Mauri2021}. This should be contrasted with the conventional Heisenberg universality class, where the critical two-point function is purely diagonal in the internal indices $\langle \phi_a(\qq) \phi_b(-\qq) \rangle =  \delta_{ab}/q^{2-\eta_H}$. Since the dipolar and Heisenberg classes are distinct, their anomalous dimensions need not coincide $\eta_D \neq \eta_H$, even if their numerical values are close within a given approximation.

Following \cite{KudlisPikelner,Nakayama2024}, we focus on the case $h = \infty$ and consider the transverse-field action
\begin{align}
\label{eq:transverseaction}
    S[\phi] =  \int_{\xx} \left[ \frac{1}{2} (\partial_a \phi_b)^2 + \frac{u_0}{2} \phi_a \phi_a + g_0 (\phi_a \phi_a)^2 \right], \quad  \partial_a \phi_a = 0,
\end{align}
which explicitly enforces the transversality of the order parameter field. Using the functional renormalization group, we systematically account for critical fluctuations of $\phi$ and determine the scaling potential and associated critical exponents directly in $d=3$ for the Aharony universality class.

\section{\label{sec:frg} Functional renormalization group treatment}

\subsection{\label{sec:weq} Flow equation and truncation}

The central object of the FRG is the effective average action $\Gamma_k[\phi]$,  a scale-dependent functional that captures fluctuations with momenta larger than the running infrared scale $k$. This construction is realized by supplementing the microscopic action with a regulator term $R_k$ that suppresses low-momentum modes, thereby rendering the path integral infrared finite.  By varying $k$, the functional  $\Gamma_k[\phi]$ provides a continuous interpolation between the microscopic description and the full effective action: at the ultraviolet scale $k=\Lambda$, the regulator freezes all fluctuations and  $\Gamma_{\Lambda}[\phi]$ coincides with the bare action, whereas in the infrared limit the regulator vanishes and $\Gamma_{k\to 0}[\phi]=\Gamma[\phi]$ evolves into the full effective action. The gradual unfreezing of fluctuations is encoded in the exact Wetterich flow equation \cite{berges2002,pawlowski2007, dupuis2021}
\begin{equation}
\label{eq:weq}
     \partial_t \G_k[\phi]  = \frac{1}{2}  \hat{\partial}_t \Tr{ \ln(\G_k^{(2)}[\phi]+R_k)}.
\end{equation}
Here, $t = \ln(k/\Lambda)$, and the derivative $\hat{\partial}_t$ acts exclusively on the regulator function $R_k$. The supertrace $\Tr$ extends over all  degrees of freedom. The operator $\Gamma_k^{(2)}[\phi]$  denotes the second functional derivative of the effective action $\Gamma_k[\phi]$ with respect to the fields of the model. 

Although the flow equation \eqref{eq:weq} is too complicated to be solved exactly, it provides a powerful basis for systematic nonperturbative truncation schemes that retain the essential physics of criticality. Throughout this work we employ the leading order of the derivative expansion of the effective average action supplemented by a scale-dependent field renormalization constant, i.e. the LPA$^{\prime}$ truncation
\begin{align}
\label{eq:ansatz}
    \G_k[\phi] = \int_{\xx} \left( U_k(\rho)  + \frac{1}{2} Z_k  (\partial_a \phi_b)^2+ \dots \right), \quad \partial_a \phi_a = 0.
\end{align}
The effective potential $U_k$ depends on the field only through the $O(d)$-invariant scalar $\rho=\phi_a\phi_a/2$. In a more general truncation, one allows for a field-dependent kinetic prefactor $Z_k(\rho)$ and expands it around the global running minimum $\rho_0=\rho_0(k)$ of $U_k(\rho)$, i.e. $  Z_k(\rho)=Z_k(\rho_0)+Z^{(1)}_k(\rho_0)\,(\rho-\rho_0)+Z^{(2)}_k(\rho_0)\,(\rho-\rho_0)^2+\dots$.
The LPA$^\prime$ approximation thus corresponds to keeping only the leading term $ Z_k \equiv Z_k(\rho_0)$, while the higher order terms $Z^{(1)}_k$, $Z^{(2)}_k,\dots$ generate momentum-dependent higher-point vertices. In what follows, we determine the coupled flows of $U_k(\rho)$ and $Z_k$ simultaneously within this approximation.

\subsection{\label{sec:uflow} Flow of the potential}

The flow equation for the scalar potential $U_k(\rho)$ is obtained by evaluating \cref{eq:weq} for a constant background field $\phi_a =  \sqrt{2 \rho} \, n_a$, where $n_a$ denotes the components of an arbitrary unit vector $\nn$.  This choice singles out a direction in the internal $O(d)$ space and thereby reduces the symmetry to the subgroup $O(d-1)$ leaving $\nn$ invariant. All vertex functions are then computed as functional derivatives of \cref{eq:ansatz} evaluated on this background. Since the truncation \cref{eq:ansatz} is defined with the constraint $\partial_a\phi_a=0$, these vertices inherit the corresponding transversality projectors.  Using $\delta \phi_a(\pp_1)/\delta \phi_b(\pp_2) =  (2\pi)^d\delta(\pp_1-\pp_2)  P_{ab}(\pp_1) $ one finds that the vertices take the form
\begin{align}
\label{eq:vertex}
     \G_{ab}^{(2)}(\pp_1,\pp_2) & = (2\pi)^d \delta(\pp_1+\pp_2) \tilde{\G}_{ab}^{(2)}(\pp_1),  \\
     \G_{abc}^{(3)}(\pp_1,\pp_2,\pp_3) &= (2\pi)^d \delta(\pp_1+\pp_2+\pp_3) P_{aa'}(\pp_1)P_{bb'}(\pp_2)P_{cc'}(\pp_3) V_{a'b'c'},
\end{align}
where 
\begin{align}
\label{eq:vertex1}
 \tilde{\G}_{ab}^{(2)}(\pp) &=  (Z_k p^2+U_k')P_{ab}(\pp) + 2 \rho U_k'' L_{a}(\pp)L_{b}(\pp),\\
    V_{a b c} &= (\delta_{ab} n_c +\delta_{b c} n_a +\delta_{c a} n_b) (2\rho)^{1/2} U_k'' + n_a n_b n_c (2\rho)^{3/2} U_k'''. 
\end{align}
Here $L_{a}(\pp) =   P_{ab}(\pp)n_b$ and the prime denotes differentiation with respect to $\rho$. To streamline the notation, we omit the explicit $k$-dependence of all vertex functions. The propagator  in the Wetterich equation (\ref{eq:weq}) $ G^{(2)} = (\G_k^{(2)}[\phi]+R_k)^{-1}$ can be then written in the form  
\begin{align}
    \label{eq:propagator}
     G^{(2)}_{a b}(\qq) &= g_1(\qq) P_{ab}(\qq) + g_2(\qq) L_{a}(\qq) L_b(\qq),
\end{align}
with 
\begin{align}
\label{eq:g1g2}
     g_1(\qq) = \frac{1}{\, Z_k q^2 +R_k(q) + U_k' \,},\quad
     g_2(\qq) = -\frac{ 2 \rho U_k'' g_1(\qq)}{\, Z_k q^2 +R_k(q) + U_k' +  2 \rho U_k'' \left[ 1 - ({\nn \!\cdot\! \qq})^2/q^2 \right]\,},
\end{align}
see \ref{sec:propagator}. The ``mass'' of the radial  mode, i.e. the components along the $L$-direction, acquires a nontrivial angular dependence through  $({\nn \!\cdot\!  \qq})^2$ in contrast to the standard Heisenberg model where the corresponding radial sector is isotropic. The projection of the equation (\ref{eq:weq}) onto the constant background yields
\begin{align}
\label{eq:uflow}
    \partial_t U_k  = \frac{1}{2}    \int_{\qq}  \tr \left( G^{(2)}(\qq)  \partial_t R_k(q)\right), \quad  \int_{\qq}  \equiv \int \frac{d^d q}{(2\pi)^d}. 
\end{align}
To investigate critical behavior, we search for scaling solutions of the flow equation (\ref{eq:uflow}). This is most conveniently done by rewriting the flow in terms of renormalized, dimensionless variables (denoted with an overbar), for which a fixed point corresponds to a scale-independent potential. We therefore introduce the dimensionless field and the dimensionless effective potential as 
\begin{align}
\label{eq:renorm_variable}
   \bar{\phi} =  k^{1 - d/2} Z_k^{1/2} \phi, \quad  u( \brho ) = k^{-d}  U_k(\rho). 
\end{align}
In these variables the flow equation becomes autonomous. All explicit scale dependence is removed, and the fixed point is given by the  stationary solution $u^*(\brho)$ satisfying $\partial_t u^*(\brho) = 0$. The scale dependence of $Z_k$ enters only through the anomalous dimension 
\begin{align}
\label{eq:eta}
  \eta = -\partial_t \ln Z_k.   
\end{align}
 In this form $\eta$ feeds back self-consistently into the fixed-point equation and  the stability analysis of perturbations around the fixed point. With this definition, the full scaling dimension of the field becomes $\Delta_{\phi} = (d-2+\eta)/2$.    

There is substantial freedom in choosing the regulator function $R_k$ in \cref{eq:weq}. 
 In this work, we employ both the optimized (Litim) cutoff \cite{litim2000, litim2001} and the exponential (Wetterich) regulator \cite{wetterich93}. Both regulators are diagonal in field space and include a dimensionless parameter $\alpha$, which is tuned according to the principle of minimal sensitivity (PMS) \cite{dupuis2021} to optimize the critical exponents. They are given by  $R_k(q) = Z_k q^2 r(q^2/k^2)$ with the dimensionless shape functions
\begin{align}
\label{eq:cutoff}
r_L(y) = \alpha \left( \frac{1}{y}-1\right)\Theta(1-y), \quad  r_W(y) = \frac{\alpha }{\exp(y)-1}.   
\end{align}
Here $\Theta$ stands for the Heaviside step function, which is treated as a weak limit of smooth functions and is further specified by the condition $\Theta(0)=1/2$.  The regulator provides an additional mass term for low-momentum modes with $q < k$, while leaving large-momentum modes $q > k$ unaffected. This choice of the Litim cutoff is convenient, as it enables a semi-analytical evaluation of the loop integrals appearing in the subsequent RG analysis. The analytical form of the flow equation  and the evaluation of the anomalous dimension $\eta$ will be presented below as an example for the Litim regulator with $\alpha = 1$. However, in our analysis, we have performed the full evaluation for both regulators, systematically tuning $\alpha$ to obtain the PMS-optimized values of the critical exponents, see \ref{sec:pms}.  By evaluating \cref{eq:uflow} we obtain the  flow  equation for the scaling potential, see \ref{appflow},
\begin{align}
\label{eq:uflow_dim}
    \partial_t u &= - d  u + (d - 2 + \eta)  \brho \, u' + \left(1 - \frac{\eta}{d+2}\right)  \left[   \frac{d-2}{\, 1 + u'\,}  +  \frac{ w_d(\bar{\ell})}{\,1 + u' +2 \brho u''\,}  \right],
\end{align}
with $ \bar{\ell} = 2 \brho u'' /(1 +u' +2 \brho u'')$. Physical fixed points correspond to solutions of the flow equation that are regular for all values of $\brho$, in particular at the origin $\brho=0$. Owing to the intrinsic nonlinearity of \cref{eq:uflow_dim}, this requirement imposes nontrivial constraints on the admissible boundary conditions for the fixed point potential $u^*(\brho)$
\begin{align}
\label{eq:boundary}
    - d  u^*(0) + \left(1 - \frac{\eta}{d+2}\right)  \frac{d-1}{\, 1 +{u'}^*(0)\,} =0,
\end{align}
thereby restricting the space of acceptable solutions.  In practice, several strategies can be employed to determine such nonsingular fixed-point solutions, including shooting methods in which ${u'}^*(0)$ is fine-tuned so that the solution remains regular over a wide range of $\brho$ \cite{Morris1994,Codello2012}; global numerical integration \cite{Borchardt2015,Murgana2023}; and local series expansions around a given point \cite{Aoki1998}. In the present work, we adopt the Taylor expansion of the potential around its running minimum, which is known to provide improved convergence properties and numerical stability compared to expansions around the origin. Taylor expansion allows the flow equations to be formulated directly in terms of physical couplings and provides direct access to the stability matrix and critical exponents.  We now turn to the computation of the anomalous dimension $\eta$ by deriving the FRG flow of the coupling $Z_k$, which closes the equation (\ref{eq:uflow_dim}).

\subsection{\label{sec:z} Wave-function renormalization and $\eta$ exponent}

The field renormalization constant $Z_k$ is defined from the momentum dependence of the two-point vertex function as
\begin{align}
\label{eq:Z_def}
    Z_k =  \lim_{p\to 0} \frac{\left[ \G_{ab}^{(2)}(\pp)\right]_{\bot} -  \left.\left[ \G_{ab}^{(2)}(\pp)\right]_{\bot}\right|_{\pp=0}}{p^2},
\end{align}
Here $ [\dots]_{\bot}$ denotes the coefficient of the transverse projector $P_{ab}(\pp)$, while contributions proportional to  the $L_{a}(\pp)L_{b}(\pp)$ structures are discarded.  Taking two functional derivatives of the Wetterich equation (\ref{eq:weq}) gives the flow of $\Gamma^{(2)}$. It involves the three- and four-point vertices. For the present truncation, the momentum dependence entering into $Z_k$ originates from the diagram with two three-point vertices; the four-point (tadpole) contribution is momentum independent at this order and drops out after the subtraction at $\pp=0$ in \cref{eq:Z_def}. Thus, the contribution relevant for the $p^2$-term is 
\begin{align}
\label{eq:gam2_flow}
    \partial_t  \G_{ab}^{(2)}(\pp) = -\frac{1}{2} P_{aa'}(\pp) P_{bb'}(\pp) \, \hat{\partial}_t \int_{\qq} \tr\left ( V^{a'} G^{(2)}(\qq) V^{b'} G^{(2)}(\pp+\qq)\right),
\end{align}
 where the matrix elements of the vertex $V^{a}$ are $(V^{a})_{bc} \equiv V_{a b c}$. Expanding the equation (\ref{eq:gam2_flow}) in $\pp$ up to the second order with the subsequent integration by part over the internal momentum $\qq$ leads
\begin{align}
\notag
   \left[ \partial_t \G_{ab}^{(2)}(\pp)\right]_{\bot} -  \left.\left[ \partial_t \G_{ab}^{(2)}(\pp)\right]_{\bot}\right|_{\pp=0}  = -\Bigg[\left(\frac{p_i p_j}{2}  \right)P_{aa'}(\pp) P_{bb'}(\pp)  \int_{\qq} &\tr\left ( 2 V^{a'} G^{(2)}(\qq)   \frac{\partial}{\partial q_i} G^{(2)}(\qq) \, V^{b'}   \frac{\partial}{\partial q_j} G^{(2)}(\qq) \partial_t R(q) \right. \\
        & \left.  +  V^{a'} \left[G^{(2)}(\qq)\right]^2\frac{\partial}{\partial q_i} \partial_t R(q) V^{b'}   \frac{\partial}{\partial q_j} G^{(2)}(\qq)\right)\Bigg]_{\bot}. 
        \label{eq:deltagamma2}
\end{align}
Substituting the explicit propagator, \cref{eq:propagator}, and the vertex, \cref{eq:vertex}, into \cref{eq:deltagamma2} and evaluating at $\rho=\rho_0$, one finds that the flow of the $p^2$-term factorizes as
\begin{align}
\label{eq:gam2_flow_structure}
     \partial_t  Z_k \, p^2  = - \left[\sqrt{2 \rho_0} U_k''(\rho_0)\right]^2 \Bigg[P_{aa'}(\pp) P_{bb'}(\pp)  \int_{\qq} &  F_{a'b'}(\pp,\qq) \Bigg]_{\bot}.  
\end{align}
The integrand $F$ is decomposed into a set of tensor structures
\begin{align}
\label{eq:F}
   F_{a'b'}(\pp,\qq) = \delta_{a'b'} (\pp \!\cdot\!\qq)^2 f_1 + q_{a'} q_{b'} (\pp\!\cdot\!\qq)^2f_2 + (p_{a'} q_{b'} + p_{b'} q_{a'}  ) f_3 + [n_{a'}, n_{b'}\text{-dependent terms}].
\end{align}
The tensor contractions and index algebra required to obtain the explicit form of $F_{a'b'}(\pp,\qq)$ were carried out with the symbolic manipulation system \textsc{FORM} \cite{KUIPERS20131453}. Terms proportional to $f_3$ drop out after contraction with the transverse projectors $P_{aa'}(\pp)P_{bb'}(\pp)$. The remaining terms $f_1$ and $f_2$ are the scalar functions of $q^2$ and $(\nn\!\cdot\!\qq)^2$ and can be reduced using the angular integrals summarized in \ref{sec:tensor_structures}. This yields the replacements
\begin{align}
     &\delta_{a'b'} (\pp \!\cdot\!\qq)^2f_1 \to  \delta_{a'b'} p ^2 \frac{q^2-(\nn \!\cdot\! \qq)^2}{d-1}f_1,\quad 
      q_{a'} q_{b'} (\pp \!\cdot\!\qq)^2f_2 \to \delta_{a'b'} p ^2  \frac{\left(q^2-(\nn \!\cdot\! \qq)^2\right)^2}{d^2-1}  f_2.
\end{align}
In the first replacement, we drop the contribution proportional to $\delta_{a'b'}(\nn\!\cdot\!\pp)^2$, which would correspond to the $O(d)$ invariant operator $(\phi_a\partial_a)^2(\phi_b\phi_b)$ in the derivative expansion. This operator is not retained in the LPA$^\prime$ truncation of the effective action, \cref{eq:ansatz}.  In the second replacement, we have omitted the terms in which the free indices $a',b'$ are carried by the external vectors $\pp$ or $\nn$, since they do not contribute after the final projection onto the $P_{ab}(\pp)$ structure.

Introducing the renormalized dimensionless variables, \cref{eq:renorm_variable}, and parametrizing the loop momentum by
$y=q^2/k^2$ and $t=(\nn\!\cdot\!\qq)/q=\cos\theta$,  we obtain the dimensionless flow equation for $\ln Z_k$ in the form
\begin{align}
\label{eq:lnz_low}
  -\partial_t \ln Z_k =  \left[2u_1 u_2^2\right] \frac{d}{2} \int\limits_{0}^{\infty}y^{d/2-1} dy  \, \frac{S_{d-1}}{S_d} \, \int \limits_{0}^{\pi} \sin (\theta)^{d-2} d \theta & \left( y \frac{1-t^2}{d-1} f_1 + y^2 \frac{\left(1-t^2\right)^2}{d^2-1}  f_2\right),
\end{align}
where $u_1, u_2$ are defined in \ref{sec:taylorMethod} and the rescaling, \cref{eq:rescaling}, is performed. The general expressions for the scalar coefficient functions $ f_1$ and $ f_2$ are provided in the Supplemental Material \cite{SM}. For the Litim regulator, only the pieces listed below contribute
\begin{equation}
\label{eq:scalar}
\begin{split}
   f_1 &= \frac{2 \left(2 a'y \,(t^{2}-1)+t^{2} a  \right) A' }{y \,a^{2} \left(a +c - c  t^{2} \right)^{2}} 
,\\
   f_2 &= \frac{8 \,t^{2} \left(a +c \right) \left(a +c + c t^{2}  \right) A}{y^{3} \left(a +c- c t^{2}  \right)^{5}}-\frac{8  \,t^{2} \left(a' y \,t^{2}+t^{2} a -a' y \right) A'}{\left(a +c - c  t^{2} \right)^{4} y^{2} \left(t^{2}-1\right)}+\frac{8  \,t^{2} A'}{\left(a +c - c  t^{2} \right)^{3} y^{2} \left(t^{2}-1\right)}+\frac{2 \left(2 a' y +a \right) A'}{\left(a +c - c  t^{2} \right)^{2} y^{2} a^{2}},
\end{split}    
\end{equation}
with $ A(y) =\partial_t R_k(q)/Z_k k^2 = - y (2 y  \partial_y r(y) + \eta r(y)) $, $A'(y) = \partial_y A(y) $, $ a(y)=  (Z_k q^2 + R_k(q))/Z_k k^2  =  y(1+r(y))$, $a'(y) =  \partial_y a(y)$ and $c = 2 u_1 u_2$. For the Litim regulator $A(y)   = \left(2+\eta \,(y -1) \right) \Theta (1-y)$, $A'(y) =\eta \, \Theta(1-y)  -2 \delta(1-y)$ and $a'(y) = \Theta(y-1)$. The terms absent in \cref{eq:scalar} are precisely those proportional to $A(y)\,a'(y)\propto \Theta(1-y)\Theta(y-1)=0$.
Substituting \cref{eq:scalar} into \cref{eq:lnz_low}, performing the $y$- and $t$-integrations at $d=3$, and solving the resulting linear equation for $\eta$ yields a closed expression for the anomalous dimension of the Aharony universality class. In the notation of \ref{sec:taylorMethod} it can be written as
\begin{align}
\label{eq:eta_final}
    \eta = \frac{6 u_1 u_2^2 \ell_0 \left(\ell_0-1 \right) \left(\left(3 \ell_0^{4}-36 \ell_0^{3}-54 \ell_0^{2}+300 \ell_0 -213\right)  w_3(\ell_0) -3 \ell_0^{3}+35 \ell_0^{2}-229 \ell_0 +213\right)}{2 u_1 u_2^2 \left(\left(3 \ell_0^{5}-27 \ell_0^{4}+174 \ell_0^{3}-294 \ell_0^{2}+159 \ell_0 -15\right) w_3(\ell_0)-3 \ell_0^{4}-102 \ell_0^{3}+244 \ell_0^{2}-154 \ell_0 +15\right)-96 \ell_0^{3}},
\end{align}
where 
\begin{align}
\ell_0 = \frac{2 u_1 u_2}{1+2 u_1 u_2}, \quad    w_3(\ell_0) = \frac{1}{\sqrt{\ell_0} }\mathrm{artanh}(\sqrt{\ell_0}).
\end{align}
 This expression is more cumbersome than its counterpart for the Heisenberg class, \cref{eq:etao3}, because the transverse projector $P_{ab}(\qq)$ introduces additional momentum-dependent tensor structures in the loop integrand. Finally, the anomalous dimension, \cref{eq:eta_final}, is to be inserted into the flow equation for the effective potential, \cref{eq:uflow_dim}. This closes the LPA$^\prime$ system. The fixed-point equation is then solved with the Taylor expansion method described in \ref{sec:taylorMethod}, yielding the numerical estimates for the critical exponents discussed in the next section.

\section{ \label{sec:results} Critical exponents}

% \begin{table}
% \centering
% \setlength{\tabcolsep}{10pt}
% \renewcommand{\arraystretch}{1.6}
% \begin{tabular}{l *{3}{S[table-format=1.6]}  *{3}{S[table-format=1.6]}} 
% \hline\hline
% & {$\eta_H$} & {$\nu_H$} & {$\omega_H$} & {$\eta_D$} & {$\nu_D$} & {$\omega_D$} \\
% \hline
% LPA\cite{Polsi20}
%   & 0       &  0.7620   &  0.7020
%   & 0       & 0.7675   & 0.7355 \\
% LPA$^\prime$\cite{Boettcher15}
%   & 0.0409  & 0.7318   & 0.7496
%   & 0.0423 & 0.7355   & 0.7908 \\
% CB \cite{HENRIKSSON20231}
%   & 0.0385(13) & 0.7120(23) & 0.791(22)
%   & {}          & {}         & {} \\
% 3-loop \cite{KudlisPikelner}
%   & 0.039(9)          & 0.700(7)         & 0.762(17)
%  &  0.033(8)  & 0.700(7)  & 0.843(19)\\
% \hline\hline
% \end{tabular}
% \caption{Critical exponents $\eta$, $\nu$, and $\omega$ obtained at the LPA and LPA$^\prime$ levels for the Heisenberg  (subscript $H$) and dipolar/Aharony  (subscript $D$) universality classes.}
% \label{tab:exponents}
% \end{table}

\begin{table}
\centering
\setlength{\tabcolsep}{10pt}
\renewcommand{\arraystretch}{1.6}
\begin{tabular}{l *{3}{S[table-format=1.6]}  *{3}{S[table-format=1.6]}} 
\hline\hline
& {$\eta_H$} & {$\nu_H$} & {$\omega_H$} \\
\hline
LPA\cite{Polsi20}
  & 0       &  0.7620   &  0.7020\\
LPA$^\prime$\cite{Boettcher15}
  & 0.0409  & 0.7318   & 0.7496\\
CB \cite{HENRIKSSON20231}
  & 0.0385(13) & 0.7120(23) & 0.791(22)\\
3-loop \cite{KudlisPikelner}
  & 0.039(9)          & 0.700(7)         & 0.762(17)\\
\hline\hline
& {$\eta_D$} & {$\nu_D$} & {$\omega_D$} \\
\hline
LPA
  & 0       &  0.7683(9)   & 0.7363(7) \\
LPA$^\prime$
  & 0.037(1) &  \multicolumn{1}{c}{0.7378$^{*}$}  & 0.786(2) \\
 3-loop \cite{KudlisPikelner}
 &  0.033(8)  & 0.700(7)  & 0.843(19)\\  
\hline\hline
\end{tabular}
\caption{Critical exponents $\eta$, $\nu$, and $\omega$ obtained in the LPA and LPA$^\prime$ approximations for the Heisenberg (subscript $H$) and dipolar/Aharony (subscript $D$) universality classes. The asterisk (*) marks the PMS estimate of the corresponding exponent derived using the Wetterich regulator. The reported values are obtained by averaging over different regulators; for details, see \ref{sec:pms}.}
\label{tab:exponents}
\end{table}

The critical exponents $\eta$, $\nu$, and $\omega$ for the Heisenberg and  Aharony universality classes are summarized in \cref{tab:exponents}. While $\eta$ is obtained in closed form from the flow of the wave-function renormalization, the remaining exponents $\nu$ and $\omega$ are extracted from the eigenvalues of the stability matrix, defined and discussed in \ref{sec:stability}.  We report the FRG estimates at two levels of approximation, LPA and LPA$^\prime$, and compare them with results from other approaches: CB data for the Heisenberg fixed point and a recent three-loop RG calculation performed directly in fixed three-dimensional space. Although higher-order perturbative estimates exist for the Heisenberg case \cite{Kompaniets2017}, we deliberately restrict the perturbative comparison to the three-loop level in order to keep a consistent benchmark with the Borel-resummed three-loop expansions available for the Aharony fixed point \cite{KudlisPikelner}.

The resulting shift of the exponent set when passing from LPA to LPA$^\prime$ is systematic and reflects the expected impact of the wave-function renormalization on the RG spectrum. For the Heisenberg universality class, this step improves the overall agreement with the CB benchmarks in \cref{tab:exponents}. Using the CB and three-loop perturbative results for the Heisenberg fixed point as references, we estimate the remaining systematic uncertainty of the FRG at the LPA$^\prime$ level to be small at the level of a few percent. Since the Aharony fixed point is scale invariant but not conformally invariant, CB determinations are not available, and the presented FRG results therefore provide an independent nonperturbative estimate of the dipolar critical exponents in three dimensions.

Table~\ref{tab:exponents} shows that the Aharony and Heisenberg exponents are numerically close. The largest relative separation appears in the correction-to-scaling exponents $\omega_H$ and $\omega_D$, whereas the differences in $\eta$ and $\nu$ remain comparatively small and lie  within the estimated systematic uncertainty of the LPA$^\prime$ truncation. The comparatively large value of  $\omega_D$ indicates a faster approach to the asymptotic scaling regime and therefore a narrower preasymptotic region in which corrections to the leading power laws must be taken into account.  While both the FRG analysis and perturbative RG calculations agree qualitatively, a quantitative spread persists, most noticeably for $\omega_D$.  We attribute this method dependence to the present level of theoretical control within each approximation scheme. On the FRG side, it is mainly driven by the truncation of the effective average action and the associated regulator dependence, while on the perturbative side, it originates from the finite loop order and resummation ambiguities. By contrast, the dependence on the Taylor polynomial truncation order can be made negligible once sufficiently high orders are retained, as demonstrated in \ref{sec:stability}.

\section{ \label{sec:conclusios} Conclusions and outlook}

In this work we have studied the critical behavior of isotropic dipolar magnets governed by the nonconformal Aharony fixed point within the functional renormalization group. Using the Wetterich flow equation and a derivative expansion truncated at the LPA$^\prime$ level, we computed the anomalous dimension $\eta$ from the flow of the wave-function renormalization and extracted the remaining critical exponents from the stability matrix of the fixed-point potential. The fixed point itself was obtained by solving the closed system of flow equations for the effective potential via the Taylor expansion around the running minimum. To reduce the dependence of our results on the choice of regulator, we
employed the principle of minimal sensitivity to optimize the critical exponents. The resulting exponents for the Aharony universality class were compared with the corresponding Heisenberg values and with available results from other approaches.

Our FRG analysis shows that the Heisenberg and dipolar fixed points are numerically close in three dimensions. Within the accuracy of the present truncation, the LPA$^\prime$ results yield a consistent set of exponents for the dipolar fixed point. A comparison across methods and truncations reveals that the dominant uncertainties are systematic, originating from the truncation of the effective average action and the regulator choice rather than from the polynomial order of the Taylor expansion, which can be taken sufficiently large to make its effect negligible.

The near coincidence of the Heisenberg and dipolar critical exponents suggests that distinguishing the two universality classes purely through $\eta$, $\nu$, and $\omega$ may be challenging in practice. It is therefore useful to consider additional universal quantities that may provide a more sensitive probe of the fixed-point structure. A promising option is universal relations among nonlinear susceptibilities. They involve higher derivatives of the equation of state, or equivalently higher-order zero-momentum vertices. These quantities are often studied in the form of universal nonlinear-susceptibility ratios, also known as $R$-ratios. They were analyzed  for the Heisenberg universality class using perturbative RG \cite{Sokolov2014, SOKOLOV2017225, KUDLIS2020114881}, FRG \cite{R_ratios} and Monte Carlo methods \cite{Campostrini2002}. Since these observables probe the global shape of the scaling equation of state, they can separate the Heisenberg and dipolar universality classes more clearly than the leading critical exponents.

  In future investigations, the present analysis can be extended by considering higher-order derivative expansions or momentum-dependent vertices \cite{Wschebor2020}. Due to the transverse structure of the dipolar model, such higher-order truncations are more involved than in the standard Heisenberg case, as they require accounting for a larger set of $O(d)$-invariant operators. These improvements are expected to further refine the estimates of critical exponents and other universal quantities at the dipolar fixed point.

\section{Acknowledgements} 
We thank M.~Kompaniets for helpful discussions on fixed-dimension renormalization-group methods.  The authors are grateful to the anonymous referee for constructive suggestions.

\appendix

\section{\label{sec:propagator}Propagator in the Wetterich equation}

 Let us consider the general case of finite $h$ in the action, \cref{eq:action}.  
For a uniform background field, one finds 
\begin{align}
    \label{eq:g2}
      \tilde{\G}_{ab}^{(2)}(\qq) &= (Z_k q^2+U_k')\,\delta_{ab} +  h  q_a q_b + 2\rho U_k'' \,n_a n_b,
\end{align}
see \cref{eq:vertex}. The regulated propagator $G^{(2)}$ is obtained by inverting
\begin{align}
\label{eq:ggam}
 \left( \tilde{\G}_{ab}^{(2)}(\qq)+  R_{ab}(\qq) \right)  G_{bc}^{(2)}(\qq) = \delta_{ac}. \end{align}
We solve  this equation using the tensor decomposition
\begin{align}
    G^{(2)}_{a b}(\qq) &= g_1(\qq) P_{ab}(\qq) + g_2(\qq) L_{a}(\qq)L_{b}(\qq) + g_3(\qq)  \frac{q_a q_b}{q^2} +  g_4(\qq) n_a n_b.
\end{align}
The scalar coefficients $g_i(\qq)$ are determined by substituting this ansatz into \cref{eq:ggam} and solving the resulting linear system. One obtains 
\begin{align}
          g_1(\qq) &=\frac{1}{a_k(q)}, \quad g_3(\qq) =  \left[a_k(q) +  h q^2 \frac{a_k(q) + c_k L(\qq)^2}{a_k(q) + c_k}\right]^{-1},\\[2ex]\quad g_4(\qq)  &= -\frac{c_k}{a_k(q) + c_k} g_3(\qq), \quad g_2(\qq) = h q^2 g_1(\qq) g_4(\qq),
\end{align}
with $a_k(q) = Z_k q^2 +R_k(q) + U_k'$,  $c_k = 2 \rho U_k''$ and  $L(\qq)^2 =  L_{a}(\qq) L_a(\qq) = 1 - ({\nn \!\cdot\! \qq})^2/q^2$.

For the Heisenberg case $h=0$, the coefficient $g_2(\qq)=0$ and the regulated propagator reduces to the standard decomposition into Goldstone (transverse to $\nn$) and radial  (longitudinal) components,
\begin{align}
    G^{(2)}_{a b}(\qq) &= \frac{\delta_{ab} - n_a n_b}{ Z_k q^2 +R_k(q) + U_k'}  + \frac{n_a n_b}{ Z_k q^2 +R_k(q) + U_k' +2 \rho U_k''}.
\end{align}
For the dipolar case $h\to \infty$, longitudinal fluctuations along $\qq$ are suppressed and the propagator becomes purely transverse. Correspondingly, the coefficients multiplying the longitudinal structures vanish, $g_3(\qq) \to 0$ and $g_4(\qq) \to 0$, and the propagator takes the form used in the main text \cref{eq:propagator}, with $g_1(\qq)$ and $g_2(\qq)$ approaching their limiting expressions given in \cref{eq:g1g2}. In this limit the inversion is performed on the transverse subspace, thus the  propagator satisfies the projected identity  $\left( \tilde{\G}_{ab}^{(2)}(\qq)+  R_{ab}(\qq) \right)  G_{bc}^{(2)}(\qq) = P_{ac}(\qq)$ rather than \cref{eq:ggam}.

\section{\label{appflow} Flow for the running potential}

Combining the equation (\ref{eq:uflow}), the expression for the propagator,  \cref{eq:propagator}, and the cut-off function, \cref{eq:cutoff}, we obtain 
\begin{align}
    \partial_t U_k  &=\frac{1}{2}    \int_{\qq} \partial_t R_k(q)  \left[  \frac{d-2}{Z_k q^2+R_k(q) + U_k'}  + \frac{1}{Z_k q^2+R_k(q)+ U_k' + 2 \rho U_k''  \left[ 1 - ({\nn \!\cdot\!\qq})^2/q^2 \right]}\right],
\end{align}
with $\partial_t R_k(q)   =  Z_k q^2  \left[  \partial_t r(q) - \eta r(q)  \right]$. Since the radial-mode mass depends on the angle between the momentum ${\qq}$ and  the direction ${\nn}$, it is convenient to employ spherical coordinates with the polar angle $\theta$ defined relative to ${\nn}$, i.e. $\theta = (\widehat{\nn,\qq})$. The momentum integration becomes
\begin{align}
    \int_{\qq} =    \frac{S_{d-1}}{(2 \pi)^d} \int\limits_{0}^{\infty} {q}^{d-1} dq  \int \limits_{0}^{\pi} \sin (\theta)^{d-2} d \theta, \quad   S_d = \frac{2\pi^{d/2}}{\Gamma(d/2)}. 
\end{align}
Carrying out angular and radial integrations using the optimized Litim cutoff, \cref{eq:cutoff}, yields
\begin{align}
\label{eq:uflow_litim}
\partial_t U_k &=  \frac{2 v_d}{d}  Z_k k^{d+2} \left(1 - \frac{\eta}{d+2}\right)  \left[   \frac{d-2}{\,Z_k k^2 + U_k'\,}  +  \frac{ w_d(\ell)}{\,Z_k k^2 + U_k' +2 \rho U_k''\,}  \right],
\end{align}
where
\begin{align}
\label{eq:wd}
 v_d = \frac{S_d}{2 (2\pi)^d}, \quad  w_d(\ell) =  \frac{S_{d-1}}{S_d} \int \limits_{0}^{\pi}  \frac{ \sin (\theta)^{d-2} d \theta }{\,1 -\ell \cos(\theta)^2 \,}, \quad  \ell = \frac{2 \rho U_k''}{\,Z_k k^2 + U_k' +2 \rho U_k''\,}.
\end{align}
 Transforming to dimensionless renormalized variables $\brho$ and $u(\brho)$ according to \cref{eq:renorm_variable} leads directly to the final form of the flow equation (\ref{eq:uflow_dim}). For convenience, we additionally apply the rescaling
\begin{align}
\label{eq:rescaling}
\brho \to \frac{2 v_d}{d} \brho, \quad u \to \frac{2 v_d}{d} u,
\end{align}
which removes the overall prefactor $2 v_d/d$  in \cref{eq:uflow_litim} as well as in the expression for $\eta$, \cref{eq:lnz_low}. This rescaling does not affect the critical exponents but merely simplifies the resulting expressions.

\section{$O(N)$-symmetric model in $d$ spatial dimension}

For completeness, we also present the RG flow equation for the effective potential in the LPA$^{\prime}$ approximation for the $O(N)$-symmetric model. In terms of the dimensionless renormalized field invariant $\brho$ and the dimensionless potential $u(\brho)$ the flow reads
\begin{align}
\label{eq:uflowo3}
\partial_t u  = & - d u+ (d - 2 + \eta ) \brho u' +   \left(1 - \frac{\eta}{d+2}\right)  \left[   \frac{N-1}{1 + u'}  +  \frac{1}{1 +u' + 2 \brho u''} \right].
\end{align}
The anomalous dimension in this truncation is given by
\begin{align}
\label{eq:etao3}
    \eta =  \frac{4 u_1 u_2^2 }{(1+2 u_1 u_2)^2}, 
\end{align}
where $u_1$ and $u_2$ are the  Taylor coefficients appearing in the expansion of the potential, see \cref{eq:taylor}. In \cref{eq:uflowo3,eq:etao3}, the rescaling, \cref{eq:rescaling}, was performed.

\section{\label{sec:taylorMethod}Series expansion of the effective potential}

Let $u_1 = u_1(t)$ be the running position of the minimum defined implicitly by the stationarity condition $u'|_{\brho=u_1} = 0$. The Taylor expansion of the potential around this point takes the form
\begin{align}
\label{eq:taylor}
   u(\brho) = u_0+ \sum_{n = 2}^{N} \frac{u_n}{n!}(\brho - u_1)^{n}.
\end{align}
The coefficient $u_0$, corresponding to the potential evaluated at its minimum, decouples from the RG flow of the remaining couplings $u_n$ with $n\geq 1$ and has no impact on universal critical behavior. It is thus omitted in the following. Since the expansion point $u_1(t)$ itself flows with the RG scale, its evolution must be determined consistently. Differentiating the condition $u'|_{\brho=u_1} = 0$  with respect to the RG time $t$ yields
\begin{align}
    \partial_t u_1 = -\frac{ \partial_t u'|_{\brho=u_1}}{u_2}.
\end{align}
To determine the beta functions of the couplings $u_n$, we expand the RG flow equation for the potential, $\partial_t u(\brho)$, around the running minimum $\brho=u_1$. Writing this expansion as 
\begin{align}
     \partial_t u(\brho) = \sum_{n = 0}^{\infty} \frac{c_n}{n!}(\brho - u_1)^{n},
\end{align}
the coefficients $c_n$ are obtained by Taylor expanding the right-hand side of the flow equation, \cref{eq:uflow_dim}. This yields a closed set of RG flow equations  for the expansion coefficients $\partial_t u_n =\beta_n$, where 
\begin{align}
\beta_1  &= -\frac{c_1}{u_2},\\
\beta_n &= c_n - u_{n+1} \frac{c_1}{u_2} , \quad n = 2,\dots, N. 
\end{align} 
For truncation at order $N$, one typically sets $u_{N+1}=0$  in the last equation. At a fixed point, all $\beta$-functions vanish simultaneously, resulting in a nonlinear algebraic system that determines the fixed point $u^*\equiv \{u_1^*,\dots,u_N^* \}$. In practice, the truncation order $N$ must be taken sufficiently large to ensure numerical convergence of the critical exponents, see \cref{fig:exponents}. Increasing $N$ systematically improves the approximation, and stable results as $N$ grows provide a consistency check for the truncation scheme. 
\begin{figure}
\centering
    \includegraphics[width=0.5\textwidth]{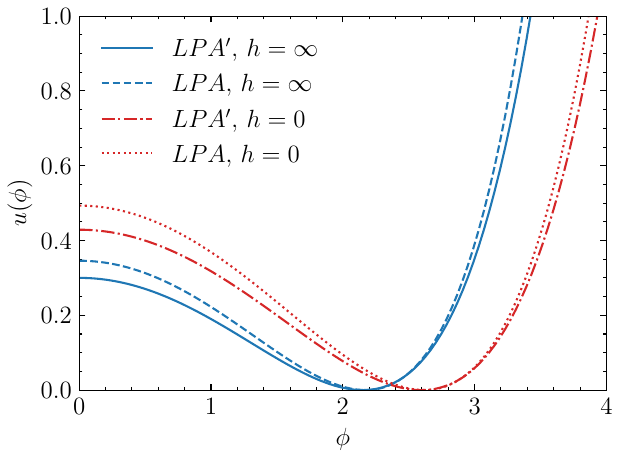}
\caption{Fixed-point potentials $u(\phi)$ at truncation order $N=16$ for the Aharony ($h=\infty$) \cref{eq:uflow_dim} and Heisenberg ($h=0$) \cref{eq:uflowo3} fixed points, shown within the LPA and LPA$^\prime$ truncations.}
    \label{fig:fixed_potential}
\end{figure}
Figure \ref{fig:fixed_potential} displays the resulting fixed-point potentials at truncation order $N=16$ for the Aharony \cref{eq:uflow_dim} and Heisenberg \cref{eq:uflowo3} cases. In each case, we plot the dimensionless fixed-point potential $u(\phi)$ reconstructed from the Taylor coefficients $u_n^*$. The two fixed points exhibit clearly distinct potential shapes, reflecting the different infrared universality classes. Comparing LPA and LPA$^\prime$, the inclusion of running wave-function renormalization produces only a mild quantitative deformation of the fixed-point potential, while leaving its qualitative structure unchanged. Although the expansion, \cref{eq:taylor}, is local around the running minimum $\brho=u_1$, the reconstructed fixed-point potential satisfies the constraint at the origin, \cref{eq:boundary}. For the truncation $N=16$ within LPA$^\prime$, we find that the residual in \cref{eq:boundary} is of order $10^{-8}$, indicating that the polynomial approximation captures the scaling solution beyond the immediate vicinity of the minimum.

\section{\label{sec:stability} The stability matrix}
To analyse the linearised RG flow in the vicinity of the fixed point $u^*$, we construct the stability matrix $\Lambda$ with the components
\begin{align}
\label{eq:stab}
    \Lambda_{i j} \equiv  \frac{\partial \beta_i}{\partial u_j} +  \frac{\partial \beta_i}{\partial \eta}  \frac{\partial \eta}{\partial u_j},  \quad  i, j=1,\dots, N,
\end{align}
where all derivatives are evaluated at the fixed point $u^*$. Since the anomalous dimension $\eta$ enters explicitly in the flow equation and is itself a function of the couplings $u_n$, its variation must also be included when linearizing the RG system. The second term in $\Lambda$ therefore accounts for the indirect contribution to the flow of $u_n$ induced by changes in $\eta$. The eigenvalues of the stability matrix characterize the scaling behavior of perturbations around the fixed-point potential. The critical exponents are directly obtained from these eigenvalues. In our convention, the correlation-length exponent $\nu$
 and the leading correction-to-scaling exponent $\omega$ are obtained from the two eigenvalues
\begin{align}
   \nu = - \frac{1}{\operatorname{Re}(\lambda_1)}, \quad \omega = \operatorname{Re}(\lambda_2), 
\end{align}
where $\lambda_1$ denotes the relevant eigenvalue with the negative real part and the largest magnitude of $\operatorname{Re}(\lambda_1)$, while $\lambda_2$ denotes the leading irrelevant one, i.e. the eigenvalue with the smallest positive real part $\operatorname{Re}(\lambda_2)$.  
\begin{figure}
    \centering  
    \includegraphics[width=\textwidth]{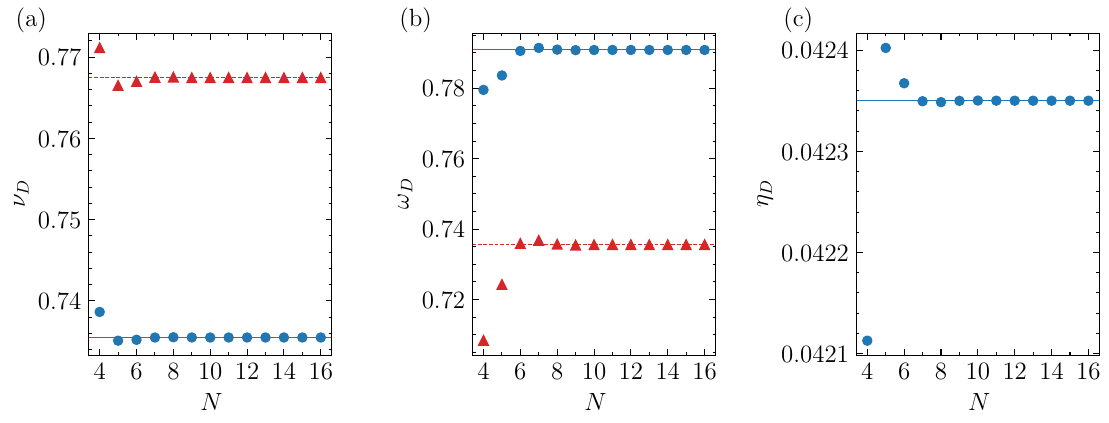}
    \caption{Convergence of the Aharony class critical exponents with the truncation order $N$. Panels show the $N$-dependence of (a) $\nu$, (b) $\omega$, and (c) $\eta$ obtained within the LPA (red triangles) and LPA$^\prime$ (blue circles) truncations.   The displayed behavior is representative for all regulator choices considered; here, results for the Litim regulator with $\alpha=1$ are shown as an example.}
    \label{fig:exponents}
\end{figure}

 In \cref{fig:exponents} we illustrate the convergence of the resulting critical exponents with the truncation order $N$. We show $\nu_D$, $\omega_D$, and $\eta_D$ obtained in both LPA and LPA$^\prime$. For sufficiently large $N$ the exponents stabilize and can be determined to arbitrarily high numerical accuracy, so the uncertainty associated with the polynomial truncation can be made negligibly small. In particular, truncating the Taylor expansion at $N=16$ is already sufficient to stabilize the results to eight decimal places for  and to six for $\omega_D$ and $\nu_D$. We therefore do not include a separate error estimate from the $N$-dependence; the accuracy is instead dominated by other sources such as the chosen FRG truncation and regulator dependence. 

\section{\label{sec:pms} Optimization of regulator dependence via the principle of minimal sensitivity}

To reduce the dependence of our results on the choice of regulator function, \cref{eq:cutoff}, we employ the principle of minimal sensitivity. The PMS asserts that a physically meaningful quantity, such as a critical exponent $X$, should be locally stationary with respect to variations of unphysical parameters, in this case the regulator parameter $\alpha$. In practice, we implement PMS by solving
\begin{equation}
\frac{\partial X(\alpha)}{\partial \alpha} = 0,
\end{equation}
where $X(\alpha)$ is the value of the exponent computed for a given regulator parameter. The stationary point $\alpha_\mathrm{PMS}$ defines the optimal choice of the regulator within the considered family, ensuring that the resulting estimate $X_\mathrm{opt} = X(\alpha_\mathrm{PMS})$
is minimally sensitive to the specific form of the regulator. This procedure has been shown to stabilize critical exponents, partially compensating for the truncation-induced regulator dependence. The residual regulator dependence is quantified by combining PMS optimization with the spread of results across different regulator families. For each regulator, the PMS condition determines an optimized value, and the central estimate $\overline{X}$ is taken as their average. The uncertainty is defined as half the difference between the maximal and minimal optimized values, $\Delta X = (X_\mathrm{opt}^\mathrm{max}- X_\mathrm{opt}^\mathrm{min})/2$, and can be expressed relative to the central value to provide a dimensionless measure of scheme dependence.  Final results in \cref{tab:exponents} are reported as the central value with a symmetric error.

\begin{table}
\centering
\setlength{\tabcolsep}{10pt}
\renewcommand{\arraystretch}{1.6}
\begin{tabular}{l *{4}{S[table-format=1.6]}  *{4}{S[table-format=1.6]}} 
\hline\hline
&  \text{regulator} & {$\eta_D$} & {$\nu_D$} & {$\omega_D$} \\
\hline
LPA
& L     & 0       & 0.76747   &  0.73555\\
& W & 0       & 0.76924   & 0.73700 \\
\hline
LPA$^\prime$
& L  & 0.03638  & \multicolumn{1}{c}{0.73550$^{*}$}  &  0.78352\\
& W & 0.03854 & 0.73782   & 0.78804 \\
\hline\hline
\end{tabular}
\caption{Critical exponents obtained via PMS within the LPA and LPA$^\prime$ truncations for the Litim (L) and Wetterich (W) regulators. $^{*}$For the Litim regulator in LPA$^\prime$, the exponent $\nu_D$ is evaluated at $\alpha=1$, since no PMS point exists.}
\label{tab:exponentsPMS}
\end{table}

\begin{figure}
    \centering  
    \includegraphics[scale=1]{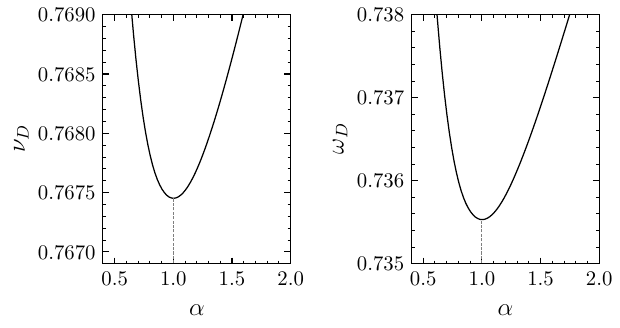}
    \caption{Critical exponents as functions of the regulator parameter $\alpha$ within the LPA truncation for the Litim regulator.}
    \label{fig:pmsLpaLit}
\end{figure}

\begin{figure}
    \centering  
    \includegraphics[scale=1]{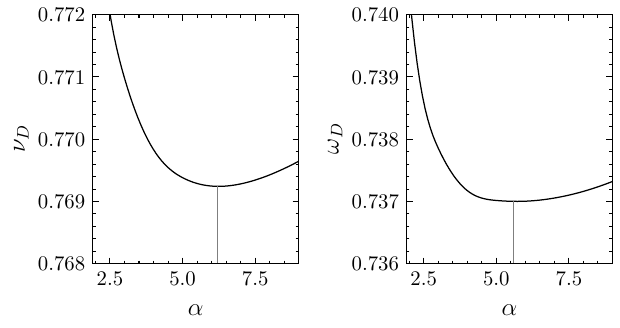}
    \caption{Critical exponents as functions of the regulator parameter $\alpha$ within the LPA truncation for the Wetterich regulator. }
    \label{fig:pmsLpaWet}
\end{figure}

\begin{figure}
    \centering  
    \includegraphics[width=\textwidth]{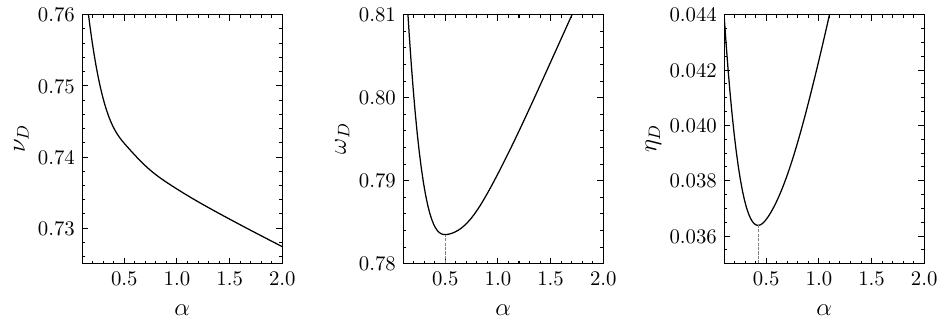}
    \caption{Critical exponents as functions of the regulator parameter $\alpha$ within the LPA$^\prime$ truncation for the Litim regulator.}
    \label{fig:pmsLpaPLit}
\end{figure}

\begin{figure}
    \centering  
    \includegraphics[width=\textwidth]{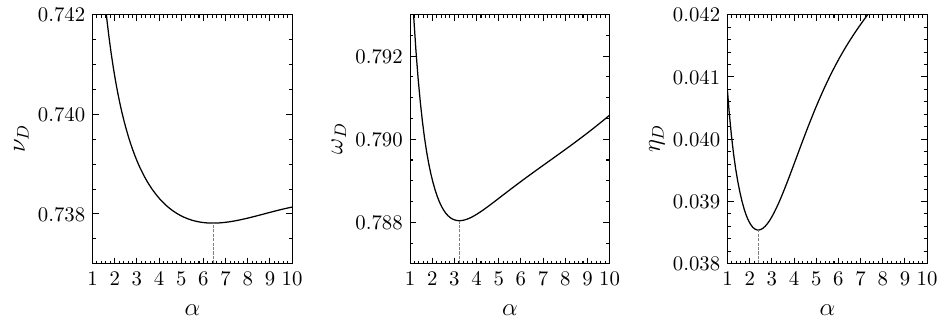}
    \caption{Critical exponents as functions of the regulator parameter $\alpha$ within the LPA$^\prime$ truncation for the Wetterich regulator. }
    \label{fig:pmsLpaPWet}
\end{figure}

Numerical results for the $\alpha$-dependence of the critical exponents, obtained within the LPA and LPA$^\prime$ truncations for both the Litim and Wetterich regulators, are shown in \cref{fig:pmsLpaLit,fig:pmsLpaPLit} and \cref{fig:pmsLpaWet,fig:pmsLpaPWet}, respectively. The corresponding PMS-optimized values are listed in \cref{tab:exponentsPMS}. For the Litim regulator within LPA$^\prime$, the PMS procedure does not produce a stationary point for $\nu_D$, and hence no optimized value can be assigned in this case, see \cref{fig:pmsLpaPLit}. This lack of an extremum can be attributed to the combined effect of the Litim regulator and the LPA$^\prime$ truncation. A similar situation was noted in \cite{Delamotte2016}, where no PMS optimum with respect to $\alpha$ was found within LPA$^\prime$ for the Wetterich regulator. In contrast, our results for the Wetterich cutoff within LPA$^\prime$, shown in \cref{fig:pmsLpaPWet}, display a clear minimum, allowing the optimized critical exponents to be defined unambiguously.

For the Litim regulator, the momentum integrals over the finite interval $ y \in [0,1]$ are evaluated using Gauss-Legendre quadrature with $10$ nodes, which is sufficient to accurately resolve the smooth behavior of the integrand on a compact domain. For the Wetterich  regulator, where the integration domain extends to $ y \in [0,\infty) $, we employ generalized Gauss-Laguerre quadrature. This choice is well adapted to the structure of the integrands, which exhibit exponential decay at large $y$ and a branch-point singularity of the form $\sqrt{y} $ at the origin. The quadrature efficiently captures both features without the need for additional regularization or domain decomposition. The convergence of the numerical integration has been carefully verified. Increasing the number of quadrature nodes beyond $10$ results in relative deviations below $0.001\%$ for all quantities considered. This demonstrates that the numerical error associated with the integration is well under control and negligible compared to the intrinsic truncation uncertainties of the LPA$^{\prime}$ approximation. After discretizing the momentum integrals via Gauss quadratures, the expression for the anomalous dimension $\eta$ becomes cumbersome. To compute its derivatives, which enter the stability matrix, \cref{eq:stab}, it is more efficient to use complex-step differentiation. In this approach, the derivative of a real function $f(x)$ is approximated as $f'(x) =\text{Im} f(x+ i \epsilon)/\epsilon + \mathcal{O}(\epsilon^2)$, where $\epsilon$ can be taken arbitrarily small, limited only by machine precision.  This approach provides very high numerical accuracy and avoids the computational cost of symbolic differentiation.

\section{\label{sec:tensor_structures} Tensor integrals}

In the presence of a fixed unit vector ${\nn}$ the momentum integrals in \cref{sec:z}  depend not only on $q^2$ but also on the product $({\nn \!\cdot\! \qq})^2$.    This appendix provides the decomposition of such integrals into the independent tensor structures allowed by the residual symmetry together with the explicit formulas for the scalar coefficient functions.

Let us consider the rank-two integrals of the form
\begin{align}
   I_{ab}[f] =  \int_{\qq} q_a q_b  f,
\end{align}
where $f= f(q^2,({\nn\!\cdot\! \qq})^2)$ is an arbitrary scalar function. The most general rank-2 tensor consistent with symmetry is a linear combination
\begin{align*}
   I_{ab}[f] = \alpha_0[f] \delta_{ab}  +  \alpha_2[f] n_a n_b.
\end{align*}
The scalar coefficients are obtained by constructing independent projections of the tensor integral, achieved by contracting indices with $\delta_{ab}$ and  $n_a n_b$. This yields
\begin{align*}
    \alpha_0[f] = \frac{1}{d-1}\int_{\qq} \left(q^2 - ({\nn \!\cdot\!\qq})^2\right) f,\quad
    \alpha_2[f] = \frac{1}{d-1}\int_{\qq} \left( ({\nn \!\cdot\!\qq})^2d - q^2 \right) f.
\end{align*}

We next consider the rank-four integral
\begin{align}
I_{abcd}[f] = \int_{\qq} \, q_a q_b q_c q_d \, f. 
\end{align}
Symmetry under permutations of indices implies the decomposition
\begin{align*}
I_{abcd}[f] = \beta_0[f] \,T^{(0)}_{abcd} + \beta_2[f] \, T^{(2)}_{abcd} + \beta_4[f] \, T^{(4)}_{abcd},
\end{align*}
via the tensor structures
\begin{align*}
T^{(0)}_{abcd} =& \delta_{ab} \delta_{cd} + \delta_{ac} \delta_{bd} + \delta_{ad} \delta_{bc}, \\
T^{(2)}_{abcd} =& \delta_{ab} n_c n_d + \delta_{ac} n_b n_d + \delta_{ad} n_b n_c  + \delta_{bc} n_a n_d + \delta_{bd} n_a n_c + \delta_{cd} n_a n_b, \\
T^{(4)}_{abcd} =& n_a n_b n_c n_d.
\end{align*}
To determine the scalar coefficients $ \beta_0[f]$, $\beta_2[f]$ and $\beta_4[f] $, it is convenient to use the scalar integrals
\begin{align*}
I_1 = \int_{\qq}  q^4 f, \quad I_2 = \int_{\qq}  q^2 ({\nn \!\cdot\! \qq})^2 f,\quad  I_3 = \int_{\qq} ({\nn \!\cdot\! \qq})^4 f.
\end{align*}
Explicit contractions  $I_{aa cc}[f]$, $I_{aa cd}[f] n_c n_d$ and $I_{abcd}[f]n_a n_b n_c n_d$ yield the linear system
\begin{align*}
I_1 =  d(d+2) \beta_0 +   2(d+2) \beta_2 + \beta_4, \quad
I_2 =  (d+2) \beta_0 +  (d+5) \beta_2 + \beta_4 ,\quad
I_3 = 3 \beta_0   + 6 \beta_2  + \beta_4.
\end{align*}
Its solution gives
\begin{align*}
 \beta_0 =  \frac{I_1 - 2 I_2 + I_3}{d^2 - 1},   \quad
\beta_2 =   \frac{(d + 3) I_2- I_1 - (d+2) I_3}{d^2 - 1},  \quad
\beta_4 =  \frac{(d^2 + 6d + 8) I_3 - 6 (d+2)I_2 + 3 I_1}{d^2 - 1}.
\end{align*}
If $f$ depends only on $q^2$, isotropy implies $\beta_0 = I_1/d(d+2)$ and $\beta_2=\beta_4=0$, consistent with the standard decomposition of a fully rotationally invariant rank-four tensor.

\bibliographystyle{elsarticle-num}
\bibliography{references}

@misc{SM,
author = {Georgii Kalagov},
  title        = {Maple File},
  year         = {2025},
  url          = {https://doi.org/10.5281/zenodo.18403539 },
  doi         = {10.5281/zenodo.18403539 },
  note = {Includes (.mw)-file with explicit expressions for the scalar coefficients used in \cref{eq:F}.}
}

@article{Delamotte2016,
  title = {Functional renormalization group approach to noncollinear magnets},
  author = {Delamotte, B. and Dudka, M. and Mouhanna, D. and Yabunaka, S.},
  journal = {Phys. Rev. B},
  volume = {93},
  issue = {6},
  pages = {064405},
  numpages = {14},
  year = {2016},
  month = {Feb},
  publisher = {American Physical Society},
  doi = {10.1103/PhysRevB.93.064405},
  url = {https://link.aps.org/doi/10.1103/PhysRevB.93.064405}
}

@article{wetterich93,
title = {Critical exponents from the effective average action},
journal = {Nuclear Physics B},
volume = {422},
number = {3},
pages = {541-592},
year = {1994},
issn = {0550-3213},
doi = {https://doi.org/10.1016/0550-3213(94)90446-4},
url = {https://www.sciencedirect.com/science/article/pii/0550321394904464},
author = {N. Tetradis and C. Wetterich}
}

@article{Boettcher15,
  title = {Scaling relations and multicritical phenomena from functional renormalization},
  author = {Boettcher, Igor},
  journal = {Phys. Rev. E},
  volume = {91},
  issue = {6},
  pages = {062112},
  numpages = {9},
  year = {2015},
  month = {Jun},
  publisher = {American Physical Society},
  doi = {10.1103/PhysRevE.91.062112},
  url = {https://link.aps.org/doi/10.1103/PhysRevE.91.062112}
}

@article{Polsi20,
  title = {Precision calculation of critical exponents in the O(N) universality classes with the nonperturbative renormalization group},
  author = {De Polsi, Gonzalo and Balog, Ivan and Tissier, Matthieu and Wschebor, Nicol\'as},
  journal = {Phys. Rev. E},
  volume = {101},
  issue = {4},
  pages = {042113},
  numpages = {22},
  year = {2020},
  month = {Apr},
  publisher = {American Physical Society},
  doi = {10.1103/PhysRevE.101.042113},
  url = {https://link.aps.org/doi/10.1103/PhysRevE.101.042113}
}

@article{Mera18,
  title = {Fast summation of divergent series and resurgent transseries from Meijer-$G$ approximants},
  author = {Mera, H\'ector and Pedersen, Thomas G. and Nikoli\ifmmode \acute{c}\else \'{c}\fi{}, Branislav K.},
  journal = {Phys. Rev. D},
  volume = {97},
  issue = {10},
  pages = {105027},
  numpages = {23},
  year = {2018},
  month = {May},
  publisher = {American Physical Society},
  doi = {10.1103/PhysRevD.97.105027},
  url = {https://link.aps.org/doi/10.1103/PhysRevD.97.105027}
}

@article{Mera16,
  title = {Hypergeometric resummation of self-consistent sunset diagrams for steady-state electron-boson quantum many-body systems out of equilibrium},
  author = {Mera, H\'ector and Pedersen, Thomas G. and Nikoli\ifmmode \acute{c}\else \'{c}\fi{}, Branislav K.},
  journal = {Phys. Rev. B},
  volume = {94},
  issue = {16},
  pages = {165429},
  numpages = {12},
  year = {2016},
  month = {Oct},
  publisher = {American Physical Society},
  doi = {10.1103/PhysRevB.94.165429},
  url = {https://link.aps.org/doi/10.1103/PhysRevB.94.165429}
}

@article{LeGuillou,
  title = {{Critical Exponents for the $n$-Vector Model in Three Dimensions from Field Theory}},
  author = {Le Guillou, J. C. and Zinn-Justin, J.},
  journal = {Phys. Rev. Lett.},
  volume = {39},
  issue = {2},
  pages = {95--98},
  numpages = {0},
  year = {1977},
  month = {Jul},
  publisher = {American Physical Society},
  doi = {10.1103/PhysRevLett.39.95},
  url = {https://link.aps.org/doi/10.1103/PhysRevLett.39.95}
}

@article{Baker,
  title = {Ising-Model Critical Indices in Three Dimensions from the Callan-Symanzik Equation},
  author = {Baker, George A. and Nickel, Bernie G. and Green, Melville S. and Meiron, Daniel I.},
  journal = {Phys. Rev. Lett.},
  volume = {36},
  issue = {23},
  pages = {1351--1354},
  numpages = {0},
  year = {1976},
  month = {Jun},
  publisher = {American Physical Society},
  doi = {10.1103/PhysRevLett.36.1351},
  url = {https://link.aps.org/doi/10.1103/PhysRevLett.36.1351}
}

@article{Aharony3,
  title = {{Critical Behavior of Magnets with Dipolar Interactions. III. Antiferromagnets}},
  author = {Aharony, Amnon},
  journal = {Phys. Rev. B},
  volume = {8},
  issue = {7},
  pages = {3349--3357},
  numpages = {0},
  year = {1973},
  month = {Oct},
  publisher = {American Physical Society},
  doi = {10.1103/PhysRevB.8.3349},
  url = {https://link.aps.org/doi/10.1103/PhysRevB.8.3349}
}

@article{Aharony2,
  title = {{Critical Behavior of Magnets with Dipolar Interactions. II. Feynman-Graph Expansion for Ferromagnets near Four Dimensions}},
  author = {Aharony, Amnon},
  journal = {Phys. Rev. B},
  volume = {8},
  issue = {7},
  pages = {3342--3348},
  numpages = {0},
  year = {1973},
  month = {Oct},
  publisher = {American Physical Society},
  doi = {10.1103/PhysRevB.8.3342},
  url = {https://link.aps.org/doi/10.1103/PhysRevB.8.3342}
}

@article{KUIPERS20131453,
title = {FORM version 4.0},
journal = {Computer Physics Communications},
volume = {184},
number = {5},
pages = {1453-1467},
year = {2013},
issn = {0010-4655},
doi = {https://doi.org/10.1016/j.cpc.2012.12.028},
url = {https://www.sciencedirect.com/science/article/pii/S0010465513000052},
author = {J. Kuipers and T. Ueda and J.A.M. Vermaseren and J. Vollinga}
}

@article{Morris1994,
title = {On truncations of the exact renormalization group},
journal = {Physics Letters B},
volume = {334},
number = {3},
pages = {355-362},
year = {1994},
issn = {0370-2693},
doi = {https://doi.org/10.1016/0370-2693(94)90700-5},
url = {https://www.sciencedirect.com/science/article/pii/0370269394907005},
author = {Tim R. Morris}
}

@article{Codello2012,
doi = {10.1088/1751-8113/45/46/465006},
url = {https://doi.org/10.1088/1751-8113/45/46/465006},
year = {2012},
month = {oct},
publisher = {IOP Publishing},
volume = {45},
number = {46},
pages = {465006},
author = {Codello, Alessandro},
title = {Scaling solutions in a continuous dimension},
journal = {Journal of Physics A: Mathematical and Theoretical}
}

@article{Borchardt2015,
  title = {Global solutions of functional fixed point equations via pseudospectral methods},
  author = {Borchardt, J. and Knorr, B.},
  journal = {Phys. Rev. D},
  volume = {91},
  issue = {10},
  pages = {105011},
  numpages = {14},
  year = {2015},
  month = {May},
  publisher = {American Physical Society},
  doi = {10.1103/PhysRevD.91.105011},
  url = {https://link.aps.org/doi/10.1103/PhysRevD.91.105011}
}

@article{Murgana2023,
  title = {{Reanalysis of critical exponents for the $O(N)$ model via a hydrodynamic approach to the functional renormalization group}},
  author = {Murgana, Fabrizio and Koenigstein, Adrian and Rischke, Dirk H.},
  journal = {Phys. Rev. D},
  volume = {108},
  issue = {11},
  pages = {116016},
  numpages = {15},
  year = {2023},
  month = {Dec},
  publisher = {American Physical Society},
  doi = {10.1103/PhysRevD.108.116016},
  url = {https://link.aps.org/doi/10.1103/PhysRevD.108.116016}
}

@article{HENRIKSSON20231,
title = {{The critical $O(N)$ CFT: Methods and conformal data}},
journal = {Physics Reports},
volume = {1002},
pages = {1-72},
year = {2023},
issn = {0370-1573},
doi = {https://doi.org/10.1016/j.physrep.2022.12.002},
url = {https://www.sciencedirect.com/science/article/pii/S0370157322004057},
author = {Johan Henriksson}
}

@article{Abhignan_2021,
   title={{Continued Functions and Perturbation Series: Simple Tools for Convergence of Diverging Series in $O(n)$-Symmetric $\phi^4$ Field Theory at Weak Coupling Limit}},
   volume={183},
   ISSN={1572-9613},
   url={http://dx.doi.org/10.1007/s10955-021-02719-z},
   DOI={10.1007/s10955-021-02719-z},
   number={1},
   journal={Journal of Statistical Physics},
   publisher={Springer Science and Business Media LLC},
   author={Abhignan, Venkat and Sankaranarayanan, R.},
   year={2021},
   month=mar }

@article{Shalaby_2021,
   title={{Critical exponents of the $O(N)$-symmetric $\phi ^4$ model from the $\varepsilon^7$ hypergeometric-Meijer resummation}},
   volume={81},
   ISSN={1434-6052},
   url={http://dx.doi.org/10.1140/epjc/s10052-021-08884-5},
   DOI={10.1140/epjc/s10052-021-08884-5},
   number={1},
   journal={The European Physical Journal C},
   publisher={Springer Science and Business Media LLC},
   author={Shalaby, Abouzeid M.},
   year={2021},
   month=jan }

@article{Mauri2021,
title = {Scale without conformal invariance in membrane theory},
journal = {Nuclear Physics B},
volume = {969},
pages = {115482},
year = {2021},
issn = {0550-3213},
doi = {https://doi.org/10.1016/j.nuclphysb.2021.115482},
url = {https://www.sciencedirect.com/science/article/pii/S0550321321001796},
author = {Achille Mauri and Mikhail I. Katsnelson}
}

@article{Nakayama2011,
title = {{What Maxwell theory in $d \neq 4$ teaches us about scale and conformal invariance}},
journal = {Nuclear Physics B},
volume = {848},
number = {3},
pages = {578-593},
year = {2011},
issn = {0550-3213},
doi = {https://doi.org/10.1016/j.nuclphysb.2011.03.008},
url = {https://www.sciencedirect.com/science/article/pii/S0550321311001465},
author = {Sheer El-Showk and Yu Nakayama and Slava Rychkov}
}

@article{NakayamaRychkov2024,
  title = {Scale without conformal invariance in dipolar ferromagnets},
  author = {Gimenez-Grau, Aleix and Nakayama, Yu and Rychkov, Slava},
  journal = {Phys. Rev. B},
  volume = {110},
  issue = {2},
  pages = {024421},
  numpages = {25},
  year = {2024},
  month = {Jul},
  publisher = {American Physical Society},
  doi = {10.1103/PhysRevB.110.024421},
  url = {https://link.aps.org/doi/10.1103/PhysRevB.110.024421}
}

@article{Nakayama2024,
  title = {Functional renormalization group approach to dipolar fixed point which is scale invariant but nonconformal},
  author = {Nakayama, Yu},
  journal = {Phys. Rev. D},
  volume = {110},
  issue = {2},
  pages = {025020},
  numpages = {7},
  year = {2024},
  month = {Jul},
  publisher = {American Physical Society},
  doi = {10.1103/PhysRevD.110.025020},
  url = {https://link.aps.org/doi/10.1103/PhysRevD.110.025020}
}

@article{KudlisPikelner,
title = {Critical behavior of isotropic systems with strong dipole-dipole interaction: Three-loop study},
journal = {Nuclear Physics B},
volume = {985},
pages = {115990},
year = {2022},
issn = {0550-3213},
doi = {https://doi.org/10.1016/j.nuclphysb.2022.115990},
url = {https://www.sciencedirect.com/science/article/pii/S0550321322003418},
author = {A. Kudlis and A. Pikelner}
}

@article{Campostrini2002,
  title = {Critical exponents and equation of state of the three-dimensional Heisenberg universality class},
  author = {Campostrini, Massimo and Hasenbusch, Martin and Pelissetto, Andrea and Rossi, Paolo and Vicari, Ettore},
  journal = {Phys. Rev. B},
  volume = {65},
  issue = {14},
  pages = {144520},
  numpages = {21},
  year = {2002},
  month = {Apr},
  publisher = {American Physical Society},
  doi = {10.1103/PhysRevB.65.144520},
  url = {https://link.aps.org/doi/10.1103/PhysRevB.65.144520}
}

@article{Sokolov2014,
  title = {{Pseudo-$\epsilon$ expansion and renormalized coupling constants at criticality}},
  author = {Sokolov, A. I. and Nikitina, M. A.},
  journal = {Phys. Rev. E},
  volume = {89},
  issue = {5},
  pages = {052127},
  numpages = {10},
  year = {2014},
  month = {May},
  publisher = {American Physical Society},
  doi = {10.1103/PhysRevE.89.052127},
  url = {https://link.aps.org/doi/10.1103/PhysRevE.89.052127}
}

@article{KUDLIS2020114881,
title = {Universal effective couplings of the three-dimensional n-vector model and field theory},
journal = {Nuclear Physics B},
volume = {950},
pages = {114881},
year = {2020},
issn = {0550-3213},
doi = {https://doi.org/10.1016/j.nuclphysb.2019.114881},
url = {https://www.sciencedirect.com/science/article/pii/S0550321319303670},
author = {A. Kudlis and A.I. Sokolov}
}

@article{SOKOLOV2017225,
title = {{Effective potential of the three-dimensional Ising model: The pseudo-$\epsilon$ expansion study}},
journal = {Nuclear Physics B},
volume = {921},
pages = {225-235},
year = {2017},
issn = {0550-3213},
doi = {https://doi.org/10.1016/j.nuclphysb.2017.05.019},
url = {https://www.sciencedirect.com/science/article/pii/S0550321317301931},
author = {A.I. Sokolov and A. Kudlis and M.A. Nikitina}
}

@article{Kompaniets2017,
  title = {{Minimally subtracted six-loop renormalization of $O(n)$-symmetric ${\ensuremath{\phi}}^{4}$ theory and critical exponents}},
  author = {Kompaniets, Mikhail V. and Panzer, Erik},
  journal = {Phys. Rev. D},
  volume = {96},
  issue = {3},
  pages = {036016},
  numpages = {26},
  year = {2017},
  month = {Aug},
  publisher = {American Physical Society},
  doi = {10.1103/PhysRevD.96.036016},
  url = {https://link.aps.org/doi/10.1103/PhysRevD.96.036016}
}

@article{Wschebor2020,
  title = {{Precision calculation of critical exponents in the $O(N)$ universality classes with the nonperturbative renormalization group}},
  author = {De Polsi, Gonzalo and Balog, Ivan and Tissier, Matthieu and Wschebor, Nicol\'as},
  journal = {Phys. Rev. E},
  volume = {101},
  issue = {4},
  pages = {042113},
  numpages = {22},
  year = {2020},
  month = {Apr},
  publisher = {American Physical Society},
  doi = {10.1103/PhysRevE.101.042113},
  url = {https://link.aps.org/doi/10.1103/PhysRevE.101.042113}
}

@Article{Aoki1998,
   author = {Ken Ichi Aoki and Keiichi Morikawa and Wataru Souma and Jun Ichi Sumi and Haruhiko Terao},
   doi = {10.1143/PTP.99.451},
   issn = {0033-068X},
   issue = {3},
   journal = {Progress of Theoretical Physics},
   month = {3},
   pages = {451-466},
   publisher = {Oxford Academic},
   title = {Rapidly Converging Truncation Scheme of the Exact Renormalization Group},
   volume = {99},
   url = {https://academic.oup.com/ptp/article/99/3/451/1845807},
   year = {1998}
}

@Article{litim2000,
title = {Optimisation of the exact renormalisation group},
journal = {Physics Letters B},
volume = {486},
number = {1},
pages = {92-99},
year = {2000},
issn = {0370-2693},
doi = {https://doi.org/10.1016/S0370-2693(00)00748-6},
url = {https://www.sciencedirect.com/science/article/pii/S0370269300007486},
author = {Daniel F. Litim}
}

@Article{litim2001,
  title = {Optimized renormalization group flows},
  author = {Litim, Daniel F.},
  journal = {Phys. Rev. D},
  volume = {64},
  issue = {10},
  pages = {105007},
  numpages = {17},
  year = {2001},
  month = {Oct},
  publisher = {American Physical Society},
  doi = {10.1103/PhysRevD.64.105007},
  url = {https://link.aps.org/doi/10.1103/PhysRevD.64.105007}
}

@Article{dupuis2021,
title = {The nonperturbative functional renormalization group and its applications},
journal = {Physics Reports},
volume = {910},
pages = {1-114},
year = {2021},
issn = {0370-1573},
doi = {https://doi.org/10.1016/j.physrep.2021.01.001},
url = {https://www.sciencedirect.com/science/article/pii/S0370157321000156},
author = {N. Dupuis and L. Canet and A. Eichhorn and W. Metzner and J.M. Pawlowski and M. Tissier and N. Wschebor}
}

@Article{berges2002,
title = {Non-perturbative renormalization flow in quantum field theory and statistical physics},
journal = {Physics Reports},
volume = {363},
number = {4},
pages = {223-386},
year = {2002},
issn = {0370-1573},
doi = {https://doi.org/10.1016/S0370-1573(01)00098-9},
url = {https://www.sciencedirect.com/science/article/pii/S0370157301000989},
author = {Jürgen Berges and Nikolaos Tetradis and Christof Wetterich},
}

@article{fisherprl,
  title = {Dipolar Interactions at Ferromagnetic Critical Points},
  author = {Fisher, Michael E. and Aharony, Amnon},
  journal = {Phys. Rev. Lett.},
  volume = {30},
  issue = {12},
  pages = {559--562},
  numpages = {0},
  year = {1973},
  month = {Mar},
  publisher = {American Physical Society},
  doi = {10.1103/PhysRevLett.30.559},
  url = {https://link.aps.org/doi/10.1103/PhysRevLett.30.559}
}

@article{dipolarI,
  title = {{Critical Behavior of Magnets with Dipolar Interactions. I. Renormalization Group near Four Dimensions}},
  author = {Aharony, Amnon and Fisher, Michael E.},
  journal = {Phys. Rev. B},
  volume = {8},
  issue = {7},
  pages = {3323--3341},
  numpages = {0},
  year = {1973},
  month = {Oct},
  publisher = {American Physical Society},
  doi = {10.1103/PhysRevB.8.3323},
  url = {https://link.aps.org/doi/10.1103/PhysRevB.8.3323}
}

@Article{pawlowski2007,
title = {Aspects of the functional renormalisation group},
journal = {Annals of Physics},
volume = {322},
number = {12},
pages = {2831-2915},
year = {2007},
issn = {0003-4916},
doi = {https://doi.org/10.1016/j.aop.2007.01.007},
url = {https://www.sciencedirect.com/science/article/pii/S0003491607000097},
author = {Jan M. Pawlowski}
}

@article{twoloops,
  title = {{Critical exponents of ferromagnets with dipolar interactions: Second-order $\ensuremath{\epsilon}$ expansion}},
  author = {Bruce, Alastair D. and Aharony, Amnon},
  journal = {Phys. Rev. B},
  volume = {10},
  issue = {5},
  pages = {2078--2087},
  numpages = {0},
  year = {1974},
  month = {Sep},
  publisher = {American Physical Society},
  doi = {10.1103/PhysRevB.10.2078},
  url = {https://link.aps.org/doi/10.1103/PhysRevB.10.2078}
}

@article{Schnetz,
  title = {Numbers and functions in quantum field theory},
  author = {Schnetz, Oliver},
  journal = {Phys. Rev. D},
  volume = {97},
  issue = {8},
  pages = {085018},
  numpages = {20},
  year = {2018},
  month = {Apr},
  publisher = {American Physical Society},
  doi = {10.1103/PhysRevD.97.085018},
  url = {https://link.aps.org/doi/10.1103/PhysRevD.97.085018}
}

@article{Kadanoff,
  title = {Static Phenomena Near Critical Points: Theory and Experiment},
  author = {KADANOFF, LEO P. and G\"OTZE, WOLFGANG and HAMBLEN, DAVID and HECHT, ROBERT and LEWIS, E. A. S. and PALCIAUSKAS, V. V. and RAYL, MARTIN and SWIFT, J. and ASPNES, DAVID and KANE, JOSEPH},
  journal = {Rev. Mod. Phys.},
  volume = {39},
  issue = {2},
  pages = {395--431},
  numpages = {0},
  year = {1967},
  month = {Apr},
  publisher = {American Physical Society},
  doi = {10.1103/RevModPhys.39.395},
  url = {https://link.aps.org/doi/10.1103/RevModPhys.39.395}
}

@article{nickel,
  title={University of Guelph report},
  author={Nickel, BG and Meiron, DI and Baker Jr, GA},
  journal={Guelph, Ontario},
  year={1977}
}

@article{ON_bootstrap,
  title={{Precision islands in the Ising and $O (N)$ models}},
  author={Kos, Filip and Poland, David and Simmons-Duffin, David and Vichi, Alessandro},
  journal={Journal of High Energy Physics},
  volume={2016},
  number={8},
  pages={1--16},
  year={2016},
  publisher={Springer},
   doi = {10.1007/JHEP08(2016)036}
}

@article{R_ratios,
  title={{Precision calculation of universal amplitude ratios in $O(N)$ universality classes: Derivative expansion results at order $O(\partial^4)$}},
  author={De Polsi, Gonzalo and Hern{\'a}ndez-Chifflet, Guzm{\'a}n and Wschebor, Nicol{\'a}s},
  journal={Physical Review E},
  volume={104},
  number={6},
  pages={064101},
  year={2021},
  publisher={APS},
  doi = {https://doi.org/10.1103/PhysRevE.104.064101},
  url = {https://journals.aps.org/pre/abstract/10.1103/PhysRevE.104.064101}
}

@article{SupressionExperiment,
  title={Depression of longitudinal fluctuations above tc of Heisenberg ferromagnets observed by polarized neutrons},
  author={K{\"o}tzler, J and G{\"o}rlitz, D and Mezei, F and Farago, B},
  journal={EPL (Europhysics Letters)},
  volume={1},
  number={12},
  pages={675--680},
  year={1986},
  doi = {10.1209/0295-5075/1/12/010},
    url = {https://iopscience.iop.org/article/10.1209/0295-5075/1/12/010/meta}
}

\end{document}